\documentclass[journal]{IEEEtran}

\usepackage[bookmarks,colorlinks]{hyperref}
\usepackage{enumitem}
\usepackage[linesnumbered,ruled,lined]{algorithm2e}

\usepackage{booktabs,array,tabularx,makecell,threeparttable}
\newcolumntype{C}[1]{>{\centering\arraybackslash}m{#1}}
\newcolumntype{Y}{>{\centering\arraybackslash}X}

\AtBeginDocument{}

\usepackage{graphicx}
\usepackage{dcolumn}
\usepackage{bm}

\usepackage[usenames,dvipsnames]{xcolor}

\usepackage{amsmath,amsfonts}
\usepackage{mathtools} 

\usepackage{array}
\usepackage[caption=false,font=normalsize,labelfont=sf,textfont=sf]{subfig}
\usepackage{textcomp}
\usepackage{stfloats}
\usepackage{url}
\usepackage{verbatim}
\usepackage{graphicx}
\usepackage{cite}
\usepackage{svg}
\usepackage[noend]{algpseudocode} 
\usepackage{etoolbox}
\usepackage{booktabs}   
\usepackage{array}      
\usepackage{multirow}

\usepackage{float}
\usepackage{pgfplots}
\usepackage[bookmarks,colorlinks]{hyperref}
\usepackage[linesnumbered,ruled,lined]{algorithm2e}
\usepackage{enumitem}
\usepackage{algpseudocode}
\usetikzlibrary{shapes.multipart,intersections}
\usepackage{cite}
\usepackage{amsmath,amssymb,amsfonts,amsthm,steinmetz}
\usepackage{graphicx}
\usepackage{mathrsfs}  
\usepackage{textcomp}
\usepackage{acronym}
\usepackage{xcolor}
\usepackage{upgreek,xspace}

\usepackage{tikz}
\usetikzlibrary{calc}
\makeatletter
\newcommand{\gettikzxy}[3]{%
  \tikz@scan@one@point\pgfutil@firstofone#1\relax
  \edef#2{\the\pgf@x}%
  \edef#3{\the\pgf@y}%
}
\makeatother

\usepackage[draft]{todonotes}

\usepackage{esvect}

\usepackage{times}
\usepackage{bm}
\usepackage{amsmath}
\usepackage{amssymb}
\usepackage{stmaryrd}
\usepackage{babel}
\usepackage{graphics, graphicx}
\usepackage{xcolor}
\usepackage{gensymb}
\usepackage{cite}
\usepackage{enumitem}
\usepackage{url}

\ifCLASSINFOpdf

\else

\fi

\usepackage[all=normal,paragraphs=normal,floats=normal,mathspacing=normal,wordspacing=normal,charwidths=tight,mathdisplays=normal,leading=normal]{savetrees}

\begin{document}

\title{Experimental Reduced-Rank Mutual Coupling Representation and Estimation for Large RIS}

\author{Philipp~del~Hougne,~\IEEEmembership{Member,~IEEE}
\thanks{This work was supported in part by the Nokia Foundation (project 20260028), the ANR France 2030 program (project ANR-22-PEFT-0005), the ANR PRCI program (project ANR-22-CE93-0010), the Rennes M\'etropole AES program (project ``SRI''), the European Union's European Regional Development Fund, and the French Region of Brittany and Rennes M\'etropole through the contrats de plan \'Etat-R\'egion program (projects ``SOPHIE/STIC \& Ondes'' and ``CyMoCoD'').}
\thanks{P.~del~Hougne is with Aalto University, Department of Electronics and Nanoengineering, 02150 Espoo, Finland and Univ Rennes, CNRS, IETR - UMR 6164, F-35000, Rennes, France (e-mail: philipp.del-hougne@univ-rennes.fr).}
}

\maketitle

\begin{abstract}
Physics-consistent optimization of reconfigurable intelligent surfaces (RISs) is thwarted in practice by the difficulty of experimentally estimating the mutual coupling (MC) between RIS elements. For large RISs, experimental MC estimation is fundamentally challenging because of the quadratic scaling of the number of unknowns with the number of RIS elements. In this Letter, we present a generic and flexible reduced-rank MC representation that allows wireless practitioners to choose a trade-off between model complexity and accuracy. We experimentally validate the direct reduced-rank MC estimation for a 100-element RIS in three radio environments (rich scattering, attenuated scattering, free space). We observe a strong environmental dependence of the influence of rank reduction on accuracy. Model-based performance evaluations highlight that the importance of MC awareness in optimization depends strongly on the radio environment and the performance indicator.

\end{abstract}


\IEEEpeerreviewmaketitle

\section{Introduction}
\label{sec_introduction}

Reconfigurable intelligent surfaces (RISs) are an emerging technology expected to play an important role in achieving the ambitious goals of next-generation wireless networks related to throughput, energy efficiency, and the integration of sensing with communications. 
From an electromagnetic perspective, an RIS is an antenna array whose ports are terminated by tunable loads; consequently, mutual coupling (MC) between the RIS elements must generally be considered. 
Under MC, the forward mapping from the RIS configuration to the corresponding end-to-end wireless channels is generally non-linear~\cite{rabault2024tacit}. 
While this non-linearity complicates the optimization of the RIS configuration, various recent works suggested that MC improves the ability to tailor the wireless channels to a desired functionality~\cite{kalde2024leveraging,wijekoon2024phase,prod2025mutual,liu2025optimization,nerini_global,semmler_decoupling,prod2025benefits}. 
Moreover, MC enables the implementation of hybrid systems capable of non-linear data transformations by encoding input data into the RIS configuration and extracting the output data from the wireless channels~\cite{momeni2023backpropagation,nerini2025analog}. 

For wireless practitioners, three important questions on MC in RISs are the following: \textit{1)} Is MC-awareness during RIS optimization important? \textit{2)}  How should MC-aware RIS optimization be implemented? \textit{3)}  How can the parameters of MC-aware system models be determined in practice?
Moreover, it is important to understand how the answers to these questions depend on (i) the targeted key performance indicator (KPI), and (ii) the hardware impairments (e.g., a limited number of states per RIS element). 
Question \textit{2)} has received significant attention in the recent theoretical literature~\cite{gradoni2021end,qian2021mutual,abrardo2021mimo,tapie2023systematic,ma2023ris,el2023optimization,wijekoon2024phase,li2024beyond,abrardo2024design,peng2025risnet,nerini_global,cheima2025}; regarding Question \textit{1)}, most of these works emphasized the importance of MC awareness. 

The main focus of this Letter is on Question \textit{3)}. Theoretical works often consider sufficiently simple settings so that the MC-aware model parameters can be evaluated analytically~\cite{gradoni2021end}. Alternatively, they can be found numerically~\cite{tapie2023systematic,zheng2024mutual,almunif2025network,macdonald2025network}. However, in realistic experiments, many system details are unknown: the radio environment is generally not known in detail; for the RIS, unknowns arise due to fabrication inaccuracies or because the design is proprietary. Moreover, even with perfect knowledge of all details, numerical evaluations would be prohibitively costly for large RISs in non-trivial radio environments. 

Experimentally, the MC-aware model parameters cannot be estimated without ambiguity because it is impossible to fulfill all the requirements outlined in~\cite{del2024virtual2p0}. Nonetheless, operationally equivalent proxy sets of parameters can be identified that accurately reproduce the mapping from RIS configuration to end-to-end wireless channels~\cite{sol2024experimentally,del2025experimental,del2025ambiguity}. However, a fundamental challenge is that the number of MC parameters scales quadratically with the number of RIS elements. In~\cite{del2025ambiguity}, an ambiguity-aware segmented estimation was proposed to avoid limitations imposed by computational resources. In~\cite{zheng2024mutual},  only accounting for coupling between nearby RIS elements and  exploiting geometric symmetries to reduce the number of MC parameters to nine was proposed. Thereby, the number of parameters was uncoupled from the number of RIS elements, but the proposed solution was limited to free-space radio environments and required a known RIS design with minimal fabrication inaccuracies. Indeed, scattering radio environments add environment-specific MC effects~\cite{rabault2024tacit} that cannot be accounted for with this approach. In~\cite{macdonald2025network}, pruning of the MC parameters was proposed, which alleviated memory and computational cost during optimization but required initially estimating all MC parameters before pruning.
Altogether, a flexible and generic approach to represent and estimate the MC parameters in a controllably reduced manner is missing.

In this Letter, we fill this gap. Our contributions are summarized as follows.
\textit{First}, we propose a flexible and generic reduced-rank MC representation suitable for direct experimental estimation. Its rank can be flexibly chosen without assuming specific geometric features; our representation smoothly interpolates between the full-rank MC-aware model and the MC-unaware model (zero-rank MC). Ultimately, this flexibility allows wireless practitioners to choose their desired trade-off between model complexity and model accuracy (as opposed to a binary choice between MC-aware and MC-unaware).
\textit{Second}, we experimentally validate our reduced-rank MC estimation for a large RIS composed of 100 1-bit-programmable, half-wavelength-sized RIS elements in three distinct radio environments: rich-scattering, strongly absorbing rich-scattering, and free space. In each case, we systematically examine the accuracy of the estimated model as a function of the chosen rank.
\textit{Third}, we perform systematic performance evaluations to examine the influence of the chosen rank on sum-rate maximization. Thereby, we determine the  trade-off between model complexity and end-to-end performance in each case.

\textit{Notation:}
$\mathbf{A} = \mathrm{diag}(\mathbf{a})$ denotes the diagonal matrix $\mathbf{A}$ whose diagonal entries are $\mathbf{a}$. 
$\mathbf{A}_\mathcal{BC}$ denotes the block of the matrix $\mathbf{A}$ whose row [column] indices are in the set $\mathcal{B}$ [$\mathcal{C}$]. 
$\mathcal{B}_i$ is the singleton containing the $i$th entry of $\mathcal{B}$.
$^\top$ and $^\dagger$ denote the transpose and transpose-conjugate operations, respectively. 
$\mathbf{I}_a$ is the $a\times a$ identity matrix.

\section{System Model}
\label{sec_system_model}

The starting point to define the  system model based on multi-port network theory (MNT) is a partition of the RIS-parametrized radio environment into three entities: \textit{i)} $N_\mathrm{A}$ antenna ports (used to inject/receive waves); \textit{ii)}  $N_\mathrm{S}$ tunable lumped elements (one per RIS element; a tunable lumped element can be described as a virtual port terminated by a tunable load); \textit{iii)} static scattering objects (including environmental scattering and structural scattering at antennas and RIS elements, all assumed to be linear, passive, time-invariant, and reciprocal).
Without assuming any specifics about geometry or material composition, we can summarize the relevant scattering characteristics of entity \textit{iii)} in a scattering matrix $\mathbf{S}\in\mathbb{C}^{N \times N}$, where $N= N_\mathrm{A} + N_\mathrm{S}$. We use the same reference impedance $Z_0$ at all ports to define $\mathbf{S}$, and we assume that the signal generators and detectors attached to the antenna ports are matched to $Z_0$. In addition, we can summarize the relevant scattering characteristics of the tunable loads in another scattering matrix $\mathbf{\Phi} = \mathrm{diag}([\rho_1, \rho_2, \dots, \rho_{N_\mathrm{S}}]) \in\mathbb{C}^{N_\mathrm{S}\times N_\mathrm{S}}$, where $\rho_i$ is the reflection coefficient of the tunable load associated with the $i$th RIS element. 

The $N_\mathrm{A}$ antennas are partitioned into $N_\mathrm{T}$ transmitting antennas and  $N_\mathrm{R}$ receiving antennas, i.e., $N_\mathrm{A}=N_\mathrm{T}+N_\mathrm{R}$. We denote by $\mathcal{T}$, $\mathcal{R}$, and $\mathcal{S}$ the sets of port indices associated with transmitting antennas, receiving antennas, and RIS elements.
According to MNT~\cite{matteo_universal,abrardo2024design}, the end-to-end wireless channel matrix $\mathbf{H}\in\mathbb{C}^{N_\mathrm{R}\times N_\mathrm{T}}$ is defined as
\begin{equation}
\mathbf{H}
= \mathbf{S}_{\mathcal{RT}}
+ \mathbf{S}_{\mathcal{RS}}\,
\bigl(\mathbf{I}_{N_\mathrm{S}}-\mathbf{\Phi}\,\mathbf{S}_{\mathcal{SS}}\bigr)^{-1}
\,\mathbf{\Phi}\,
\mathbf{S}_{\mathcal{ST}}.
\label{eq1}
\end{equation}
The matrix inversion in (\ref{eq1}) is responsible for the aforementioned non-linear mapping from RIS configuration to end-to-end wireless channels. 
Assuming $\mathbf{S}_\mathcal{SS}=\mathbf{0}$, (\ref{eq1}) simplifies to the widespread cascaded (CASC) model:
\begin{equation}
    \mathbf{H}^\mathrm{CASC} = {\mathbf{S}}_\mathcal{RT} +{\mathbf{S}}_\mathcal{RS} \mathbf{\Phi} {\mathbf{S}}_\mathcal{ST},
    \label{eq2}
\end{equation}
where the mapping from RIS configuration to end-to-end wireless channels is affine.

In practice, the wireless practitioner does not directly choose the reflection coefficient $\rho_i$ of the load terminating the $i$th RIS element's virtual port. Instead, the wireless practitioner chooses the $j$th out of $m$ available states for the load in question. The mapping from the chosen state index to the corresponding reflection coefficient is not necessarily known and/or trivial. If the RIS design is unknown (e.g., because it is proprietary) or if the utilized tunable lumped element is not sufficiently documented, the possible values of $\rho_i$ are unknown. Moreover, if the load circuit is not designed carefully enough (e.g., if the supplied power is insufficient), there may be complicated coupling mechanisms in the control signal circuit that make the control signal of one lumped element dependent on all the other elements' control signals. For simplicity, we assume that all $N_\mathrm{S}$ tunable lumped elements are identical and independently 1-bit-programmable.\footnote{How to treat more general cases is described and experimentally validated in~\cite{del2025experimental}; we deliberately avoid this more general description here because it is unnecessarily complicated for the RIS prototype utilized in this paper.} Thus, there are only two possible loads with reflection coefficients $\rho_-$ and $\rho_+$; we do \textit{not} assume  these reflection coefficients to be known. The reflection coefficient of the load terminating the virtual port associated with the $i$th RIS element is then 
\begin{equation}
    \rho_i = \rho_- + b_i (\rho_+-\rho_-),
\end{equation}
where $b_i\in \{0,1\}$ is the binary control signal for the tunable lumped element associated with the $i$th RIS element. The control vector of the RIS configuration is thus $\mathbf{b}=[b_1, b_2,\dots,b_{N_\mathrm{S}}]\in\{0,1\}^{N_\mathrm{S}}$ and 
\begin{equation}
    \mathbf{\Phi}(\mathbf{b}) = \rho_-\mathbf{I}_{N_\mathrm{S}}+ (\rho_+-\rho_-) \mathrm{diag}(\mathbf{b}).
    \label{eq_phi_b}
\end{equation}

To summarize, the wireless practitioner can choose $\mathbf{b}$ and measure the corresponding $\mathbf{H}$. The required parameters to map $\mathbf{b}$ to $\mathbf{H}$ using the MNT system model based on (\ref{eq_phi_b}) and (\ref{eq1}) are $\rho_-$ and $\rho_+$ as well as the entries of $\mathbf{S}_\mathcal{RT}$, $\mathbf{S}_\mathcal{RS}$, $\mathbf{S}_\mathcal{SS}$ (which is known to be symmetric due to reciprocity), and $\mathbf{S}_\mathcal{ST}$.

\section{Reduced-Rank MC Representation in MNT}
\label{sec_principle}

As detailed in Sec.~\ref{sec_ParamEstimAlg} and~\cite{del2025experimental,del2025ambiguity}, an operationally equivalent proxy for $ \mathbf{S}_{\mathcal{RT}}$ can be identified based on a single measurement, and a matching operational proxy for $ \mathbf{S}_{\mathcal{RS}}$ and $ \mathbf{S}_{\mathcal{ST}}$ can be identified based on $N_\mathrm{S}$ additional measurements. The challenge in experimentally estimating the MNT model parameters lies in estimating a matching operational proxy for the RIS elements' MC matrix $\mathbf{S}_{\mathcal{SS}}$. The number of complex-valued unknowns in $\mathbf{S}_{\mathcal{SS}}$ is $N_\mathrm{S}(N_\mathrm{S}+1)/2$ under our assumption of reciprocity (and $N_\mathrm{S}^2$ otherwise), i.e., it scales quadratically with $N_\mathrm{S}$. 

\newcommand{\EK}{\mathbf{E}_k}
\newcommand{\ei}[1]{\mathbf{e}_{#1}}

Sparser representations of $\mathbf{S}_{\mathcal{SS}}$ reduce the number of unknown parameters. As mentioned,~\cite{zheng2024mutual,macdonald2025network} explored sparsity in the canonical basis but relied on the absence of rich scattering in the radio environment and required prior knowledge about the RIS design. The methods from~\cite{zheng2024mutual,macdonald2025network} also do not apply to RISs composed of arbitrarily distributed RIS elements (even if they are deployed in free space). 
Inspired by dimensionality reductions in other areas of electromagnetism~\cite{brick2016fast,brick2018rapid,fromenteze2021lowering}, we propose a reduced-rank representation of $\mathbf{S}_{\mathcal{SS}}$. Being a complex-symmetric matrix, $\mathbf{S}_{\mathcal{SS}}$ admits a Takagi decomposition:
\begin{equation}
    \mathbf{S}_{\mathcal{SS}} = \mathbf{U}\mathbf{\Sigma}\mathbf{U}^\top,
\end{equation}
where $\mathbf{U}\in\mathbb{C}^{N_{\mathrm S}\times N_{\mathrm S}}$ is unitary and collects the singular vectors, and $\mathbf{\Sigma}=\mathrm{diag}([\sigma_1,\ldots,\sigma_{N_{\mathrm S}}])$ is diagonal and collects the non-negative real singular values $\sigma_i$. We assume that the singular values are ordered: $\sigma_1\geq\sigma_2\geq\dots\geq\sigma_{N_\mathrm{S}}$.
For our rank-$k$ approximation $\widehat{\mathbf{S}}_{\mathcal{SS}}(k)$ of $\mathbf{S}_\mathcal{SS}$, we retain only the $k$ dominant singular values and vectors. With $\EK = \bigl[\ei{1},\dots,\ei{k}\bigr] \in \mathbb{R}^{N_{\mathrm S}\times k}$, where $\ei{j}\in\mathbb{R}^{N_{\mathrm S}}$ is the $j$th canonical basis vector, we define $ \mathbf{U}_k = \mathbf{U}\EK$ and $ \mathbf{\Sigma}_k = \EK^{\top}\mathbf{\Sigma}\EK$, yielding
\begin{equation}
\widehat{\mathbf{S}}_{\mathcal{SS}}(k) = \mathbf{U}_k\mathbf{\Sigma}_k\mathbf{U}_k^\top.
\label{eq:sss-trunc}
\end{equation}
The number of real-valued degrees of freedom (DOFs) in $\widehat{\mathbf{S}}_{\mathcal{SS}}(k)$ is $2N_\mathrm{S}k - k(k-1)$:
$\mathbf{\Sigma}_k$ has $k$ real-valued DOFs; $\mathbf{U}_k$ has $2N_\mathrm{S}k-k^2$ real-valued DOFs.\footnote{$\mathbf{U}_k$ has $N_\mathrm{S}k$ complex-valued entries. The constraint $\mathbf{U}_k^\dagger\mathbf{U}_k=\mathbf{I}_k$ removes $k^2$ real DOFs: $k$ real-valued constraints for the diagonal entries and $k(k-1)/2$ complex-valued constraints (i.e., $k(k-1)$ real-valued constraints) for the off-diagonal entries. Thus, $\mathbf{U}_k$ has $2N_\mathrm{S}k-k^2$ real-valued DOFs.}

There are two important limits of the reduced-rank MC representation in (\ref{eq:sss-trunc}): $k=N_{\mathrm S}$ recovers the full MNT model with $N_\mathrm{S}(N_\mathrm{S}+1)$ real-valued DOFs for $\mathbf{S}_\mathcal{SS}$, while $k=0$ yields the MC-unaware CASC model with zero DOFs for $\mathbf{S}_\mathcal{SS}$. In between these limits, (\ref{eq:sss-trunc}) provides wireless practitioners with a discrete ``rank-knob'' $k$ that yields a monotonic trade-off between accuracy and complexity as $k$ increases. 

From the point of view of parameter estimation, our rank-$k$ approximation of $\mathbf{S}_\mathcal{SS}$ reduces the number of real-valued unknown parameters in $\mathbf{S}_\mathcal{SS}$ from $N_{\mathrm S}(N_{\mathrm S}+1)$ to $2N_\mathrm{S}k-k(k-1)$. To map our $2N_\mathrm{S}k-k(k-1)$ real-valued unknown parameters to $\widehat{\mathbf{S}}_{\mathcal{SS}}(k)$, we use the following factorization:
\begin{equation}
\widehat{\mathbf{S}}_{\mathcal{SS}}(k)
=
\begin{bmatrix}
\mathbf I_k & \mathbf A^\top
\end{bmatrix}^{\top}
\mathbf B
\begin{bmatrix}
\mathbf I_k & \mathbf A^\top
\end{bmatrix},
\label{eq5}
\end{equation}
where $\mathbf{A}\in\mathbb{C}^{(N_\mathrm{S}-k)\times k}$ and $\mathbf{B}=\mathbf{B}^\top\in\mathbb{C}^{k \times k}$. The number of parameters matches exactly the one identified for (\ref{eq:sss-trunc}): $\mathbf{A}$ has $2(N_\mathrm{S}-k)k$ real-valued DOFs and $\mathbf{B}$ has $k(k+1)$ real-valued DOFs. However, (\ref{eq5}) is easier to use in parameter estimation because it avoids the need to enforce the column-orthonormality constraints on $\mathbf{U}_k$. (\ref{eq:sss-trunc}) and (\ref{eq5}) are generally equally expressive; the only assumption in (\ref{eq5}) is that $\mathbf{E}_k^\top\widehat{\mathbf{S}}_{\mathcal{SS}}(k)\mathbf{E}_k$ is invertible. This assumption is generally valid because only carefully tuned or symmetric systems feature the reflectionless scattering modes implied by vanishing singular values of $\mathbf{E}_k^\top\widehat{\mathbf{S}}_{\mathcal{SS}}(k)\mathbf{E}_k$~\cite{sweeney2020theory}.

\section{MNT Parameter Estimation Algorithm}
\label{sec_ParamEstimAlg}

Our MNT parameter estimation algorithm seeks to identify a set of operationally equivalent proxy parameters allowing us to accurately map $\mathbf{b}$ to $\mathbf{H}$ using (\ref{eq_phi_b}) and (\ref{eq1}). We decorate our proxy parameters with a tilde to distinguish them from the ground-truth parameters.
Our first two steps follow~\cite{del2025experimental}; the subsequent (main) step differs from~\cite{del2025experimental} because it makes use of our reduced-rank representation.

\textit{Step 1:} We define $\Tilde{\rho}_- \triangleq 0$. Moreover, we define an RIS reference configuration $\mathbf{b}_-$ in which all  RIS elements are in their ``$-$'' state, i.e., $\rho_i = \rho_- \ \forall \ i$. It follows from (\ref{eq1}) that $\Tilde{\mathbf{S}}_\mathcal{RT}\triangleq \mathbf{H}(\mathbf{b}_-)$.

\textit{Step 2:} We define $\mathbf{b}_i$ as the RIS configuration in which all RIS elements are in their ``$-$'' state except for the $i$th one which is in its ``$+$'' state. For each $1\leq i\leq N_\mathrm{S}$, we measure $\mathbf{H}(\mathbf{b}_i)$, evaluate $\mathbf{\Delta}_i = \mathbf{H}(\mathbf{b}_i)-\mathbf{H}(\mathbf{b}_-)$, compute the singular value decomposition (SVD) $\mathbf{\Delta}_i = \mathbf{U}_i\mathbf{\Sigma}_i\mathbf{V}_i^\dagger$, and identify the first left singular vector $\mathbf{u}_i$ and the first right singular vector $\mathbf{v}_i$.
We define $\tilde{\mathbf{S}}_{\mathcal{RS}_i} \triangleq \alpha_i \mathbf{u}_i$ and $\tilde{\mathbf{S}}_{\mathcal{S}_i\mathcal{T}} \triangleq \beta_i \mathbf{v}_i^\dagger$, where $\alpha_i$ and $\beta_i$ are complex-valued scaling factors that we determine in Step 3. 

\textit{Step 3:} We measure $\mathbf{H}(\mathbf{b}^{(m)})$ for $M$ random (but known) choices of $\mathbf{b}^{(m)}$, yielding $M$ matching pairs: $\{\mathbf{b}^{(m)},\mathbf{H}(\mathbf{b}^{(m)})\}_{m=1}^M$. We collect all real-valued unknowns in $\bm\theta \in\mathbb{R}^u$, where $u=4N_\mathrm{S}+2(N_\mathrm{S}-k)k+k(k+1)+2$. The four terms in the definition of $u$ correspond, respectively, to the scaling factors $\alpha_i$ and $\beta_i$, the entries of $\mathbf{A}$, the entries of $\mathbf{B}$, and $\rho_+$. We define the cost function to be minimized as
\begin{equation}
\mathcal{L}(\bm\theta) = \frac{\displaystyle\sum_{m=1}^{M_\mathrm{b}}\bigl\|\,{\mathbf H}^\mathrm{PRED}(\mathbf b^{(m)};\boldsymbol\theta)
-\mathbf H^\mathrm{MEAS}(\mathbf{b}^{(m)})\,\bigr\|_{1}}
{\displaystyle\sum_{m=1}^{M_\mathrm{b}}\bigl\|\,\mathbf H^\mathrm{MEAS}(\mathbf{b}^{(m)})\,\bigr\|_{1}},
\end{equation}
where $\|\cdot\|_{1}$ denotes the element-wise sum of magnitudes over all complex-valued entries, and $M_\mathrm{b}$ denotes the batch size. The superscripts MEAS and PRED indicate, respectively, whether the channel matrix is measured or predicted based on our reduced-rank MNT model with parameters $\bm\theta$.
We split our $M$ pairs $\{\mathbf{b}^{(m)},\mathbf{H}(\mathbf{b}^{(m)})\}$ into $M_\mathrm{val}=100$ pairs for validation and $M_\mathrm{train}=M-M_\mathrm{val}$ pairs for training. Using the Adam optimizer with a decaying step size, we optimize $\bm\theta$ to minimize $\mathcal{L}(\bm\theta)$. We use the full batch of $M_\mathrm{train}$ training examples in every iteration. We monitor the validation loss and stop training when it plateaus. We retain the set of parameters corresponding to the lowest validation loss. We repeat this procedure five times with different random initializations, and retain the set of parameters that yielded the smallest validation loss. 

Because our reduced-rank MNT model specializes to CASC with $k=0$, we do not need a separate algorithm for CASC.

\section{Experimental Validation and \\Performance Analysis}

\subsection{Experimental Setups}
\label{subsec_ExpSetup}

Our experiments are based on an RIS prototype comprising 225 1-bit-programmable, half-wavelength-sized RIS elements of which we use 100 in this Letter; the remaining ones are kept in their reference state throughout all experiments. RIS design details are discussed in~\cite{ahmed2025over}. The utilized tunable lumped elements (PIN diodes) are point-like compared to the wavelength at our operating frequency of 2.45~GHz, ensuring the validity of our system model~\cite{del2025ambiguity}. We use four parallel, half-wavelength-spaced antennas (ANT-W63WS2-SMA) as transmit and receive arrays.
We consider the three distinct radio environments shown in Fig.~\ref{Fig1}: \textit{A)} a rich-scattering reverberation chamber (RS-RC; as in~\cite{del2025ambiguity}), \textit{B)} an attenuated reverberation chamber (A-RC), and \textit{C)} an anechoic chamber (AC). The mode stirrer in the RC is static throughout all experiments.

For a given RIS configuration $\mathbf{b}$, we directly measure the corresponding $\mathbf{H}(\mathbf{b})$ using an eight-port vector network analyzer. 

\subsection{Experimental Reduced-Rank MNT Parameter Estimation}
\label{subsec_ExpParamEst}

\begin{figure*}
\centering
\includegraphics[width=\textwidth]{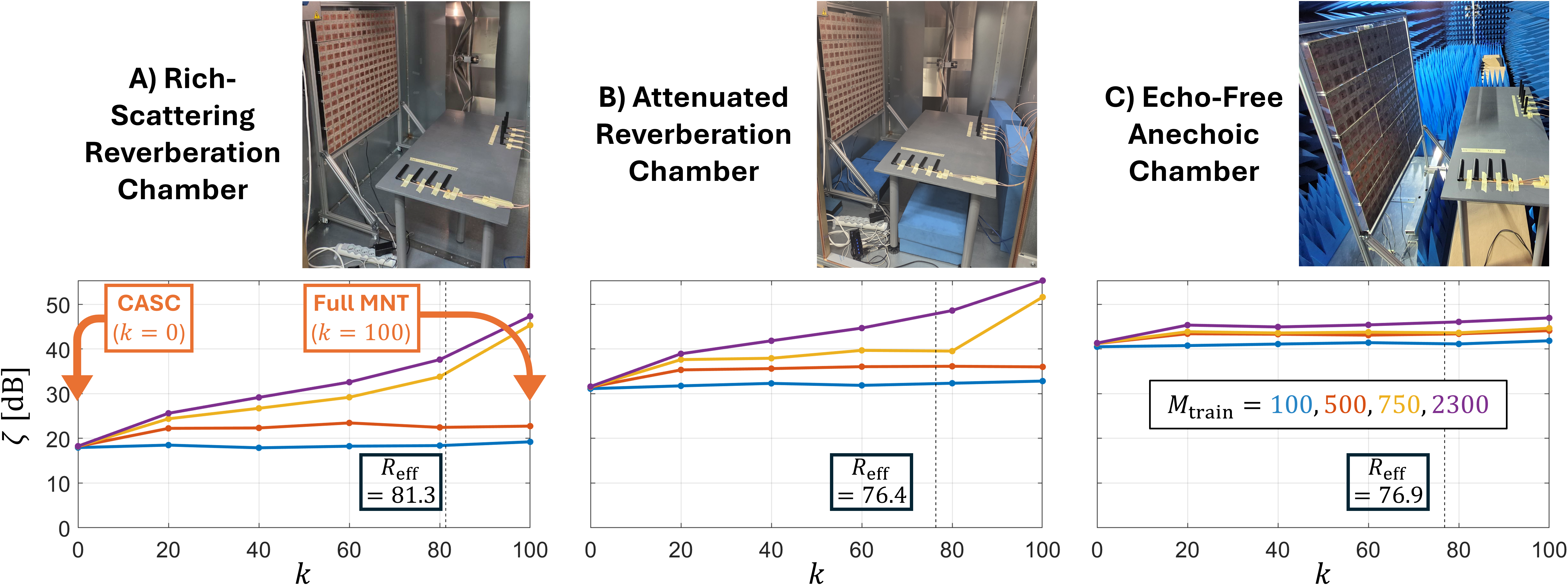}
\caption{Achieved model accuracies $\zeta$ (evaluated with $M_\mathrm{test}=100$ unseen RIS configurations) as a function of the radio environments (distinct panels), $k$ (horizontal axis) and $M_\mathrm{train}$ (colors). The values of $R_\mathrm{eff}$ (indicated by a vertical dashed line) refer to the effective rank of $\tilde{\mathbf{S}}_\mathcal{SS}$ estimated with $k=N_\mathrm{S}=100$ and $M_\mathrm{train}=2300$. $k=0$ and $k=N_\mathrm{S}=100$ correspond, respectively, to CASC and the full MNT model. }
\label{Fig1}
\end{figure*}

To quantify the model accuracy, we evaluate the following metric on a test batch of $M_\mathrm{test}=100$ unseen random RIS configurations:
\begin{equation}
\zeta = \mathrm{SD} \big[h_{ij}^{\mathrm{GT}}(\mathbf{b})\big] \ \big/ \ \mathrm{SD} \big[h_{ij}^{\mathrm{GT}}(\mathbf{b}) - h_{ij}^{\mathrm{PRED}}(\mathbf{b})\big],
    \label{eq_zeta}
\end{equation}
where SD denotes the standard deviation across all channel coefficients and considered RIS configurations, $h_{ij}$ is the $(i,j)$th channel coefficient, and GT and PRED refer, respectively, to measured ground truth and model prediction.

Our results obtained with the algorithm from Sec.~\ref{sec_ParamEstimAlg} are displayed for the three radio environments and different choices of $M_\mathrm{train}$ as a function of $k$ in Fig.~\ref{Fig1}. In all three radio environments, we achieve $\zeta>45~\mathrm{dB}$ with the full MNT model ($k=N_\mathrm{S}=100$) and $M_\mathrm{train}=2300$. However, the influence of MC awareness on $\zeta$ drastically varies between the three environments. With the CASC model ($k=0$), we achieve 41.3~dB in the AC, 31.6~dB in the A-RC, and only 18.2~dB in the RS-RC. This observation emphasizes that MC between RIS elements is \textit{not} fixed solely by the RIS design; it also depends significantly on the radio environment~\cite{rabault2024tacit}. Interestingly, the achieved values of $\zeta$ for CASC display no significant dependence on $M_\mathrm{train}$ (within the explored range); this makes sense because the number of real-valued unknowns is $u_\mathrm{CASC} = u(k=0) = 4N_\mathrm{S}+2$ and each of our measurements yields $2N_\mathrm{R}N_\mathrm{T}$ independent real-valued constraints so that $2N_\mathrm{R}N_\mathrm{T}M_\mathrm{train}>u_\mathrm{CASC}$ for all considered values of $M_\mathrm{train}$.

In the RS-RC, we observe (in line with~\cite{sol2024experimentally}) that $\zeta$ increases rather abruptly as a function of $M_\mathrm{train}$ when $M_\mathrm{train}$ crosses a threshold: with $k=100$, increasing $M_\mathrm{train}$ from 100 to 500 barely improves $\zeta$ from 19.2 to 22.7, but increasing $M_\mathrm{train}$ further to 750 yields a massive accuracy jump to $\zeta=45.3\ \mathrm{dB}$; further increasing $M_\mathrm{train}$ to 2300 yields only a marginal further improvement to 47.3~dB. Below the threshold, $\zeta$ displays no substantial dependence on $k$, whereas above the threshold increasing $k$ results in a clearly higher value of $\zeta$. 

In the A-RC, the trends are qualitatively similar but less pronounced than in the RS-RC because MC is notably weaker. In the AC, MC is so weak that no significant dependence of $\zeta$ on $k$ or $M_\mathrm{train}$ is observable.

Interestingly, the effective rank $R_\mathrm{eff}(\tilde{\mathbf{S}}_\mathcal{SS}) = \mathrm{exp} \left(-\sum_{i=1}^{N_\mathrm{S}} \hat{\sigma}_i \mathrm{ln}(\hat{\sigma}_i)\right)$~\cite{roy2007effective}, where $\hat{\sigma}_i=\sigma_i / \sum_{l=1}^{N_\mathrm{S}} {\sigma_l}$, is similarly high in all three radio environments for $k=N_\mathrm{S}=100$ and $M_\mathrm{train}=2300$. Our interpretation is that while a coarser approximation of $\tilde{\mathbf{S}}_\mathcal{SS}$ results in a lower loss of model accuracy when MC is weaker, this does not imply substantial changes of the diversity of $\tilde{\mathbf{S}}_\mathcal{SS}$.

\subsection{Performance Evaluation}
\label{subsec_PerfEval}

We limit our performance evaluation to the goal of maximizing the sum rate in the RS-RC because the sum rate was the most MC-sensitive KPI in~\cite{del2025ambiguity} and the RS-RC is the only one of our three setups with substantial MC. In the A-RC and the AC, we found that the weak influence of MC on the model accuracy translates into almost identical KPIs for all considered values of $k$ and $M_\mathrm{train}$. We define the sum rate as in [Sec.~V.A,~\cite{del2025ambiguity}] and use the coordinate-descent algorithm for RIS optimization defined in [Sec.~V.B,~\cite{del2025ambiguity}]. We use the Woodbury identity to expedite the algorithm's model evaluations~\cite{prod2023efficient}.

Our results are summarized in Table~\ref{Table1}. While the full MNT model ($k=N_\mathrm{S}=100$) with $M_\mathrm{train}=2300$ achieves a 7.2-fold improvement of the sum-rate compared to the use of a random RIS configuration, it requires $u = 10502$ real-valued parameters. Meanwhile, our reduced-order models with $k=40$ and $k=20$ only require $u=6842$ and $u=4022$ real-valued parameters and still achieve a 5.5-fold and 4.9-fold improvement of the sum rate, respectively. The CASC model (with $k=0$) involves $u=402$ real-valued parameters and achieves a 4.1-fold sum-rate improvement. Overall, $k$  thus offers a convenient tuning knob to wireless practitioners to trade off model complexity with end-to-end performance.

\begin{table}
\centering
\caption{Predicted and achieved sum-rate in bits/s/Hz as a function of $k$ and $M_\mathrm{train}$. The model for $k=N_\mathrm{S}=100$ and $M_\mathrm{train}=2300$ with $\zeta = 47.3\ \mathrm{dB}$ (see Fig.~\ref{Fig1}) serves as ground truth (GT). The sum rate with a random RIS configuration is $1.99\pm0.38$~bits/s/Hz.  }
\begin{tabular}{
    ||
    >{\centering\arraybackslash}p{0.6cm}||
    >{\centering\arraybackslash}p{0.5cm}|
    >{\centering\arraybackslash}p{0.5cm}||
    >{\centering\arraybackslash}p{0.5cm}|
    >{\centering\arraybackslash}p{0.5cm}||
    >{\centering\arraybackslash}p{0.5cm}| 
    >{\centering\arraybackslash}p{0.5cm}||
    >{\centering\arraybackslash}p{0.5cm}|
    >{\centering\arraybackslash}p{0.5cm}||
}
\hline
\multirow{2}{*}{\!\!\!$M_\mathrm{train}$} & 
\multicolumn{2}{c||}{$k=0$} & 
\multicolumn{2}{c||}{$k=20$} & 
\multicolumn{2}{c||}{$k=40$} & 
\multicolumn{2}{c||}{$k=100$} \\ 
\cline{2-9}
 & GT & \!\!\textcolor{gray}{PRED} & GT & \!\!\textcolor{gray}{PRED} & GT & \!\!\textcolor{gray}{PRED} & GT & \!\!\textcolor{gray}{PRED}  \\
\hline\hline
100  & 7.93 & \textcolor{gray}{7.31} & 6.53 & \!\!\textcolor{gray}{10.17} & 5.88 & \!\!\textcolor{gray}{10.57} & 6.28 & \!\!\textcolor{gray}{10.43}  \\
500   & 8.12 & \textcolor{gray}{7.19} & 7.63 & \!\!\textcolor{gray}{10.63} & 8.26 & \!\!\textcolor{gray}{15.13} & 9.29 & \!\!\textcolor{gray}{12.81}  \\
750   & 7.93 & \textcolor{gray}{7.12} & \!\!11.46 & \!\!\textcolor{gray}{16.44} & \!\!10.82  & \!\!\textcolor{gray}{13.38} & \!\!12.34 & \!\!\textcolor{gray}{12.48}  \\
2300   & 8.11 & \textcolor{gray}{7.10} & 9.66 & \!\!\textcolor{gray}{11.94} & \!\!10.96 & \!\!\textcolor{gray}{11.96} & \!\!14.26 & \!\!\textcolor{gray}{14.26}  \\
\hline
\end{tabular}
\label{Table1}
\end{table}

\section{Conclusion}

We have introduced a generic and flexible reduced-rank representation of MC between RIS elements that can be directly estimated experimentally without any assumptions about the RIS design or radio environment. Our representation enables a monotonic trade-off between accuracy and complexity; its limiting cases are the MC-unaware CASC model and the full MNT model. We experimentally verified our approach in three radio environments with a 100-element RIS, systematically studying how the MC rank trades off model complexity for model accuracy and end-to-end performance.  

\section*{Acknowledgment}
The author acknowledges I.~Ahmed, F. Boutet, and C. Guitton, who, under the author's supervision, previously built the RIS prototype for the work presented in~\cite{ahmed2025over}. The author further acknowledges J.~Sol, who provided technical support for setting up the experiments.

\bibliographystyle{IEEEtran}

\begin{thebibliography}{10}
\providecommand{\url}[1]{#1}
\csname url@samestyle\endcsname
\providecommand{\newblock}{\relax}
\providecommand{\bibinfo}[2]{#2}
\providecommand{\BIBentrySTDinterwordspacing}{\spaceskip=0pt\relax}
\providecommand{\BIBentryALTinterwordstretchfactor}{4}
\providecommand{\BIBentryALTinterwordspacing}{\spaceskip=\fontdimen2\font plus
\BIBentryALTinterwordstretchfactor\fontdimen3\font minus
  \fontdimen4\font\relax}
\providecommand{\BIBforeignlanguage}[2]{{%
\expandafter\ifx\csname l@#1\endcsname\relax
\typeout{** WARNING: IEEEtran.bst: No hyphenation pattern has been}%
\typeout{** loaded for the language `#1'. Using the pattern for}%
\typeout{** the default language instead.}%
\else
\language=\csname l@#1\endcsname
\fi
#2}}
\providecommand{\BIBdecl}{\relax}
\BIBdecl

\bibitem{rabault2024tacit}
A.~Rabault \emph{et~al.}, ``On the tacit linearity assumption in common
  cascaded models of {RIS}-parametrized wireless channels,'' \emph{IEEE Trans.
  Wirel. Commun.}, vol.~23, no.~8, pp. 10\,001--10\,014, 2024.

\bibitem{kalde2024leveraging}
J.~Kalde \emph{et~al.}, ``Leveraging mutual coupling in antenna-amplifier
  systems for increased reconfigurability,'' \emph{IEEE J. Microw.}, 2024.

\bibitem{wijekoon2024phase}
D.~Wijekoon \emph{et~al.}, ``Phase shifter optimization in {RIS}-aided {MIMO}
  systems under multiple reflections,'' \emph{IEEE Trans. Wirel. Commun.},
  vol.~23, no.~8, pp. 8969--8983, 2024.

\bibitem{prod2025mutual}
H.~Prod'homme and P.~del Hougne, ``Mutual coupling in dynamic metasurface
  antennas: Foe, but also friend,'' \emph{IEEE Wirel. Commun.}, vol.~32, no.~4,
  pp. 30--36, 2025.

\bibitem{liu2025optimization}
Y.~Liu and C.~D. Sarris, ``Optimization of reconfigurable intelligent surface
  codebooks using a mutual coupling aware {CNN} model of scattered fields,''
  \emph{IEEE Antennas Wirel. Propag. Lett.}, vol.~24, no.~9, pp. 3188--3192,
  2025.

\bibitem{nerini_global}
M.~Nerini \emph{et~al.}, ``Global optimal closed-form solutions for intelligent
  surfaces with mutual coupling: Is mutual coupling detrimental or
  beneficial?'' \emph{IEEE Trans. Wirel. Commun.}, 2025.

\bibitem{semmler_decoupling}
D.~Semmler \emph{et~al.}, ``Decoupling networks and super-quadratic gains for
  {RIS} systems with mutual coupling,'' \emph{IEEE Trans. Wirel. Commun.},
  2025.

\bibitem{prod2025benefits}
H.~Prod'homme \emph{et~al.}, ``Benefits of mutual coupling in dynamic
  metasurface antennas,'' \emph{arXiv:2502.15565}, 2025.

\bibitem{momeni2023backpropagation}
A.~Momeni \emph{et~al.}, ``Backpropagation-free training of deep physical
  neural networks,'' \emph{Science}, vol. 382, no. 6676, pp. 1297--1303, 2023.

\bibitem{nerini2025analog}
M.~Nerini and B.~Clerckx, ``Analog computing for signal processing and
  communications -- part {I}: Computing with microwave networks,''
  \emph{arXiv:2504.06790}, 2025.

\bibitem{gradoni2021end}
G.~Gradoni and M.~Di~Renzo, ``End-to-end mutual coupling aware communication
  model for reconfigurable intelligent surfaces: An electromagnetic-compliant
  approach based on mutual impedances,'' \emph{IEEE Wirel. Commun. Lett.},
  vol.~10, no.~5, pp. 938--942, 2021.

\bibitem{qian2021mutual}
X.~Qian and M.~Di~Renzo, ``Mutual coupling and unit cell aware optimization for
  reconfigurable intelligent surfaces,'' \emph{IEEE Wirel. Commun. Lett.},
  vol.~10, no.~6, pp. 1183--1187, 2021.

\bibitem{abrardo2021mimo}
A.~Abrardo \emph{et~al.}, ``{MIMO} interference channels assisted by
  reconfigurable intelligent surfaces: Mutual coupling aware sum-rate
  optimization based on a mutual impedance channel model,'' \emph{IEEE Wirel.
  Commun. Lett.}, vol.~10, no.~12, pp. 2624--2628, 2021.

\bibitem{tapie2023systematic}
J.~Tapie \emph{et~al.}, ``Systematic physics-compliant analysis of over-the-air
  channel equalization in {RIS}-parametrized wireless networks-on-chip,''
  \emph{IEEE J. Sel. Areas Commun.}, vol.~42, no.~8, pp. 2026--2038, 2024.

\bibitem{ma2023ris}
R.~Ma \emph{et~al.}, ``{RIS}-assisted {SWIPT} network for internet of
  everything under the electromagnetics-based communication model,'' \emph{IEEE
  Internet Things J.}, vol.~11, no.~9, pp. 15\,402--15\,415, 2024.

\bibitem{el2023optimization}
H.~El~Hassani \emph{et~al.}, ``Optimization of {RIS}-aided {MIMO}—a mutually
  coupled loaded wire dipole model,'' \emph{IEEE Wireless Commun. Lett.},
  vol.~13, no.~3, pp. 726--730, 2023.

\bibitem{li2024beyond}
H.~Li \emph{et~al.}, ``Beyond diagonal reconfigurable intelligent surfaces with
  mutual coupling: Modeling and optimization,'' \emph{IEEE Commun. Lett.},
  vol.~28, no.~4, pp. 937--941, 2024.

\bibitem{abrardo2024design}
A.~Abrardo \emph{et~al.}, ``Design of reconfigurable intelligent surfaces by
  using {S}-parameter multiport network theory—optimization and full-wave
  validation,'' \emph{IEEE Trans. Wireless Commun.}, vol.~23, no.~11, pp.
  17\,084--–17\,102, 2024.

\bibitem{peng2025risnet}
B.~Peng \emph{et~al.}, ``{RISnet}: A domain-knowledge driven neural network
  architecture for {RIS} optimization with mutual coupling and partial {CSI},''
  \emph{Trans. Wireless. Comm.}, vol.~24, no.~5, p. 4469–4482, Feb. 2025.

\bibitem{cheima2025}
C.~Hammami \emph{et~al.}, ``Statistical multiport-network modeling and
  efficient discrete optimization of {RIS},'' \emph{arXiv:2508.01776}, 2025.

\bibitem{zheng2024mutual}
P.~Zheng \emph{et~al.}, ``Mutual coupling in {RIS}-aided communication: Model
  training and experimental validation,'' \emph{IEEE Trans. Wirel. Commun.},
  vol.~23, no.~11, pp. 17\,174--17\,188, 2024.

\bibitem{almunif2025network}
M.~Almunif \emph{et~al.}, ``Network-based design of reactive beamforming
  metasurfaces,'' \emph{IEEE Trans. Antennas Propag.}, vol.~73, no.~8, pp.
  5970--5980, 2025.

\bibitem{macdonald2025network}
T.~I. MacDonald \emph{et~al.}, ``Network based hybrid spatial-spectral {RIS}
  synthesis method,'' \emph{Proc. EuCAP}, pp. 01--04, 2025.

\bibitem{del2024virtual2p0}
P.~del Hougne, ``Virtual {VNA 2.0}: Ambiguity-free scattering matrix estimation
  by terminating not-directly-accessible ports with tunable and coupled
  loads,'' \emph{IEEE Trans. Antennas Propag.}, vol.~73, no.~7, pp. 4903--4908,
  2025.

\bibitem{sol2024experimentally}
J.~Sol \emph{et~al.}, ``Experimentally realized physical-model-based frugal
  wave control in metasurface-programmable complex media,'' \emph{Nat.
  Commun.}, vol.~15, no.~1, p. 2841, 2024.

\bibitem{del2025experimental}
P.~del Hougne, ``Experimental multiport-network parameter estimation and
  optimization for multi-bit {RIS},'' \emph{IEEE Wirel. Commun. Lett.}, 2025.

\bibitem{del2025ambiguity}
------, ``Ambiguity-aware segmented estimation of mutual coupling in large
  {RIS}: Algorithm and experimental validation,'' \emph{arXiv:2507.22750},
  2025.

\bibitem{matteo_universal}
M.~Nerini \emph{et~al.}, ``A universal framework for multiport network analysis
  of reconfigurable intelligent surfaces,'' \emph{IEEE Trans. Wirel. Commun.},
  vol.~23, no.~10, pp. 14\,575--14\,590, 2024.

\bibitem{brick2016fast}
Y.~Brick and A.~E. Y{\i}lmaz, ``Fast multilevel computation of low-rank
  representation of {$\mathcal{H}$}-matrix blocks,'' \emph{IEEE Trans. Antennas
  Propag.}, vol.~64, no.~12, pp. 5326--5334, 2016.

\bibitem{brick2018rapid}
------, ``Rapid rank estimation and low-rank approximation of impedance matrix
  blocks using proxy grids,'' \emph{IEEE Trans. Antennas Propag.}, vol.~66,
  no.~10, pp. 5359--5369, 2018.

\bibitem{fromenteze2021lowering}
T.~Fromenteze \emph{et~al.}, ``Lowering latency and processing burden in
  computational imaging through dimensionality reduction of the sensing
  matrix,'' \emph{Sci. Rep.}, vol.~11, no.~1, p. 3545, 2021.

\bibitem{sweeney2020theory}
W.~R. Sweeney \emph{et~al.}, ``Theory of reflectionless scattering modes,''
  \emph{Phys. Rev. A}, vol. 102, no.~6, p. 063511, 2020.

\bibitem{ahmed2025over}
I.~Ahmed \emph{et~al.}, ``Over-the-air emulation of electronically adjustable
  {Rician} {MIMO} channels in a programmable-metasurface-stirred reverberation
  chamber,'' \emph{IEEE Trans. Antennas Propag.}, vol.~73, no.~4, pp.
  2104--2119, 2025.

\bibitem{roy2007effective}
O.~Roy and M.~Vetterli, ``The effective rank: A measure of effective
  dimensionality,'' \emph{Proc. EUSIPCO}, pp. 606--610, 2007.

\bibitem{prod2023efficient}
H.~Prod’homme and P.~del Hougne, ``Efficient computation of physics-compliant
  channel realizations for (rich-scattering) {RIS}-parametrized radio
  environments,'' \emph{IEEE Commun. Lett.}, vol.~27, no.~12, pp. 3375--3379,
  2023.

\end{thebibliography}

\providecommand{\noopsort}[1]{}\providecommand{\singleletter}[1]{#1}%

\end{document}